# The Effect of Ageing on the Structure and Properties of Model Liquid Infused Surfaces

*Sarah J. Goodband[1], Steven Armstrong[2], Halim Kusumaatmaja[1]\* and Kislon Voitchovsky[1†]*

1. Department of Physics, Durham University, Durham DH1 3LE, UK
2. Smart Materials & Surfaces Laboratory, Faculty of Engineering & Environment, Northumbria University, Newcastle upon Tyne NE18ST, U.K.

## Abstract

Liquid infused surfaces (LIS) exhibit unique properties that make them ideal candidates for a wide range of applications, from anti-fouling and anti-icing coatings to self-healing surfaces and controlled wetting. However, when exposed to realistic environmental conditions, LIS tend to age and progressively lose their desirable properties, potentially compromising their application. The associated ageing mechanisms are still poorly understood, and results reflecting real-life applications are scarce. Here we track the ageing of model LIS composed of glass surfaces functionalized with hydrophobic nanoparticles and infused with silicone oil. The LIS are fully submerged in aqueous solutions and exposed to acoustic pressure waves for set time intervals. The ageing is monitored by periodic measurements of the LIS' wetting properties. We also track changes to the LIS' nanoscale structure. We find that the LIS rapidly lose their slippery properties due to a combination of oil loss, smoothing of the nanoporous functional layer and substrate degradation when directly exposed to the solution. The oil loss is consistent with water microdroplets entering the oil layer and displacing oil away from the surface. These mechanisms are general and could play a role in the ageing of most LIS.

## Introduction

Liquid infused surfaces (LIS) represent a family of functional surfaces inspired by the nepenthes pitcher plant whose porous outer surface is imbibed with a lubricating liquid. This effectively replaces the plant's exposed outer surface with a fluid layer that creates a slippery surface able to shed liquid droplets and trap insects.[1] LIS are of considerable economical interest because they provide a non-toxic method for preventing the fouling and corrosion of surfaces by blocking the attachment of organisms or blocking direct interaction between the solid support and the outside environment.[2,3] Potential applications range from reducing the natural fouling of buildings, windows, transport vehicles and underwater structures (e.g. rigs, turbines, water treatment systems and power plants) to preventing biofilm formation on the surface of medical devices and implants.[4–7] Moreover, LIS have been shown to be anti-icing[8–10] and self-healing,[2,11] exhibit low roll off angles[12] and drag reduction,[13] present a high optical transparency[2,9] and may be used for fog-harvesting applications.[14]

Regardless of intended application, all designs of LIS have to meet three main criteria: (1) the chemical affinity between the lubricating fluid and the solid should be higher than that between the ambient fluid and the solid, (2) the solid should preferably be roughened so as to increase the surface area for the adhesion of the lubricating fluid and its immobilization, and (3) the lubricating fluid and the ambient fluid must be largely immiscible.[2,3] Since the initial development of LIS,[2,15–17] many different geometries and materials have been proposed for the porous substructure and lubricating fluid. These include the development of flexible surfaces from self-assembling polymers[18] or using novel ferro-fluids to infuse surfaces.[19] Experimental advances have further been complemented by theoretical studies to examine ideal geometries and the interplay of the infused liquid with supported liquid droplets moving across LIS.[20–22]

While there is abundant research investigating the design of LIS, their ageing and wear is often overlooked despite it being considered one of the biggest problems facing self-cleaning surfaces.[23] When available, studies typically examine LIS ageing during static storage in air or in solution at room temperature,[10] or under steady external perturbations on freshly made LIS.[3,24–26] Durability is assessed in terms of substrate recovery after damage incidents (such as incision or impact in the infused layer),[4,5,27] oil loss, and the ageing of surfaces under soaking conditions. Generally, LIS retain their slippery properties[8,28,29] provided the oil layer is not depleted, and periodic re-immersion in oil has been shown to allow most LIS to regain their self-healing properties and high droplet mobility.[30] Indeed, in nature, the nepenthes pitcher plant exhibits a unique system of continuous liquid transport which is used to allow the surface to retains its slippery properties.[31] Synthetic mimics have also been designed to replicate this spreading behavior.[32–34]

This suggests the integrity of the infused oil layer to be the single most important factor; upon oil depletion LIS progressively lose their anti-fouling properties and bio-material is able to attach.[28] Several mechanisms may be responsible for the oil loss, including exposure to shear flows,[24–26] failure under gravity,[26] and aqueous droplet cloaking by the oil resulting in LIS material being carried away.[35]

In this study we investigate the ageing of model LIS exposed to an environment that aims to mimic real-life applications such as waves in the sea, rain falling or localized impacts. This is achieved by immersing our model LIS in aqueous solutions and exposing them to ultrasonic pressure waves. The use of well-defined ultrasonic waves ensures a reproducible but accelerated LIS ageing compared to ambient laboratory conditions. This strategy enables us to identify some of the mechanisms responsible for oil loss and degradation of the porous layer, including the impact of dissolved salt ions in the aqueous solution that are in contact with the LIS. Significantly, we track the functional evolution of the ageing LIS and link it to nanoscale changes that occur within the different LIS components. We do this by combining macroscopic contact angle (CA) and contact angle hysteresis (CAH) measurements with atomic force microscopy (AFM) of the porous and liquid-infused surface. This approach allows for a systematic study into the effects of ageing across different length scales.

## Experimental Methods

**Preparation of LIS substrates**

Glass slides were prepared following a literature protocol as described elsewhere.[36] Briefly, glass substrates were cleaned using Decon 90 (Sigma-Aldrich-Merck, Gillingham, UK), followed by alternating steps of rinsing and sonication (30 min bursts) in ultrapure water (18.2 MΩ, Merck-Millipore, Hertfordshire, UK). Slides were then left to dry in air. Subsequent rinsing of the slides was carried out consecutively in acetone (purity 99% (Emplura©), Sigma-Aaldrich-Merck, Gillingham, UK), and isopropanol (purity 99.8 %, Fisher Scientific, Loughbrough, UK) and dried under a stream of nitrogen. After 30 mins in air, a layer of hydrophobized nanoparticles was sprayed evenly across the slide surface (GLACO™ spray, SOFT 99 Corp. Japan) and left to dry for 60 mins. Additional layers were applied every hour until a total of five coats was achieved unless otherwise specified. A drop (0.5 mL) of silicone oil (20 cSt @ 25 °C, Sigma-Aaldrich-Merck, Gillingham, UK) was then placed at the center of the slides and immediately spin coated (1000 rpm 1 min, then 500 rpm 1 min). Slides were used fresh, and any storage (outside ageing) was done with the slides placed in petri-dishes with closed lids at ambient temperatures.

**Preparation of dichlorodimethylsilane hydrophobized glass substrate**

Glass slides were hydrophobized with dichlorodimethylsilane (DMS) to serve in control experiments. The preparation followed established protocols.[37] Slides were soaked sequentially in acetone and isopropanol, each for a minimum of 30 mins. They were then dried using a stream of nitrogen, plasma cleaned for 15 mins (>30 W, VacuLAB-X, Tantec, UK) and subsequently dehydrated in an oven at 100 °C for 60 mins. The slides were then immediately placed inside a desiccator next to 1 ml of DMS placed in an open dish. The desiccator was then placed under vacuum overnight to allow for DMS vapor deposition on the slides. After functionalization, the slides were rinsed with acetone and ultrapure water and dried overnight at 40 °C.

**Ageing using static soaking**

Freshly prepared nanoparticle-functionalized slides and the LIS were placed in a sealed beaker containing either ultrapure water or a 600 mM NaCl solution. The samples were removed periodically to make contact angle measurements or for nanoscale imaging with AFM.

**Accelerated ageing using sonication**

Samples were placed in a beaker containing either ultrapure water or a 600 mM NaCl solution and sonicated in bursts of 1 min, using a VWR, USC-TH bath sonicator (VWR, Lutterworth, UK). The ultrasonic bath operates at 45 kHz and has an average output power of 180 W. Using a bespoke liquid displacement sensor built from piezo-ceramic bi-morph (RS PRO Vibration Sensor, model 285-784, RS Components, Northants, NN17 9RS, UK), it was possible to estimate the average ultrasonication power at the location of the sample, yielding a value of $800 \pm 400$ $Wm^{-2}$. The associated oscillatory displacement velocity of the aqueous solution at the sample's surface is in the order of 8 $ms^{-1}$. Details about the setup geometry and the deduction of the above estimations are presented in the supporting online information (section 1 and Fig. S1). Similar to static ageing, samples were removed periodically to make contact angle measurements or for nanoscale imaging with AFM.

**High-resolution AFM imaging**

In order to fit into the AFM chamber, glass slides were cut into small (<10 mm) pieces and epoxy-glued to steel disks (12 mm, SPI Supplies, West Chester, USA) before undergoing the ageing procedures described above (either sonication or soaking). AFM imaging was carried out in amplitude-modulation using a commercial Cypher ES equipped with photothermal excitation of the cantilever (Asylum Research, Oxford Instruments, Santa Barbara CA, USA). Imaging was carried out in air or in the aqueous solutions using Arrow-UFHAuD-10 cantilevers (nominal stiffness of 1 N/m, Nanoworld, Neuchatel, Switzerland). Image optimization was achieved following established protocols.[38]

**Contact angle and contact angle hysteresis measurements**

All contact angle (CA) images were captured using a portable digital microscope (Dino-lite Edge) and analyzed using the ImageJ freeware, in particular FIJI plugins Dropsnake[39,40] (100 µl drops) or ContactAngle[41] (10-20 µl drops). Static CA measurements were conducted with 100 µl droplets. For contact angle hysteresis (CAH) measurements, a 10 µl droplet was first deposited on the surface. A second 10 µl droplet was then added to measure the advancing angle and subsequently removed to obtain the receding angle while video-recording the experiment. Stills of the videos were then analyzed to infer the advancing and receding angles.

**Characterization of the infused oil layer thickness**

The nanoparticle coated slides were weighed before and after spin coating with silicone oil. Knowing the oil density and the area of the slide, a thickness for the oil film could be derived. This approach is convenient to track oil losses and rapidly estimate changes in the infused layer thickness. However, it rests on the assumption of a uniform, homogenous oil film and neglects any of the oil contained within the rough porous nanoparticle layer. These assumptions are justified considering the order magnitude in thickness difference between the oil and porous layers, at least for fresh samples. Values obtained are in the range of 6 – 8 µm. Alternatively, an estimate of the oil film thickness can be derived from the equation for thin film thickness after spin coating. This determines the oil film thickness to be around 5 µm, in good agreement with weight measurements (see section 2 of the Supporting Information).

## Results and Discussion

In order to objectively and systematically investigate the ageing of model LIS surfaces, we have organized the study in three consecutive parts. First, we establish the ideal LIS model to be used in the ageing study and characterize its surface properties at the nano and macroscopic scale. The stability of the resulting LIS is also examined (standard ageing under nominal laboratory conditions and with no external perturbations). Second, we investigate the impact of ultrasonication and soaking on the LIS and their nanoparticle (NP)-based retention substrates, both from a structural and functional perspective. Finally, the different experiments are brought together, revealing a dynamical ageing mechanism able to remove the infused LIS liquid and degrade the underlying nanoporous NP substrate.

### *Determination of an Ideal LIS Model*

To maximize the practical relevance of our findings, we use a model LIS made with a commercially available spray of hydrophobized silica nanoparticles (GLACO$^{TM}$, see Materials and Methods section) which creates a porous nano-layer that can be readily infused with silicone oil.[42] GLACO$^{TM}$ -based LIS are often used to create inexpensive and facile LIS model systems,[36,43,44] making them an ideal research platform for this study. There is however, no set protocol to create an optimal porous layer of hydrophobized nanoparticles (NPs). We therefore started by testing a variety of NPs-functionalized support with a varying number of NP layers for oil infusion. We characterized these supports and tested the resulting LIS (Fig. 1).

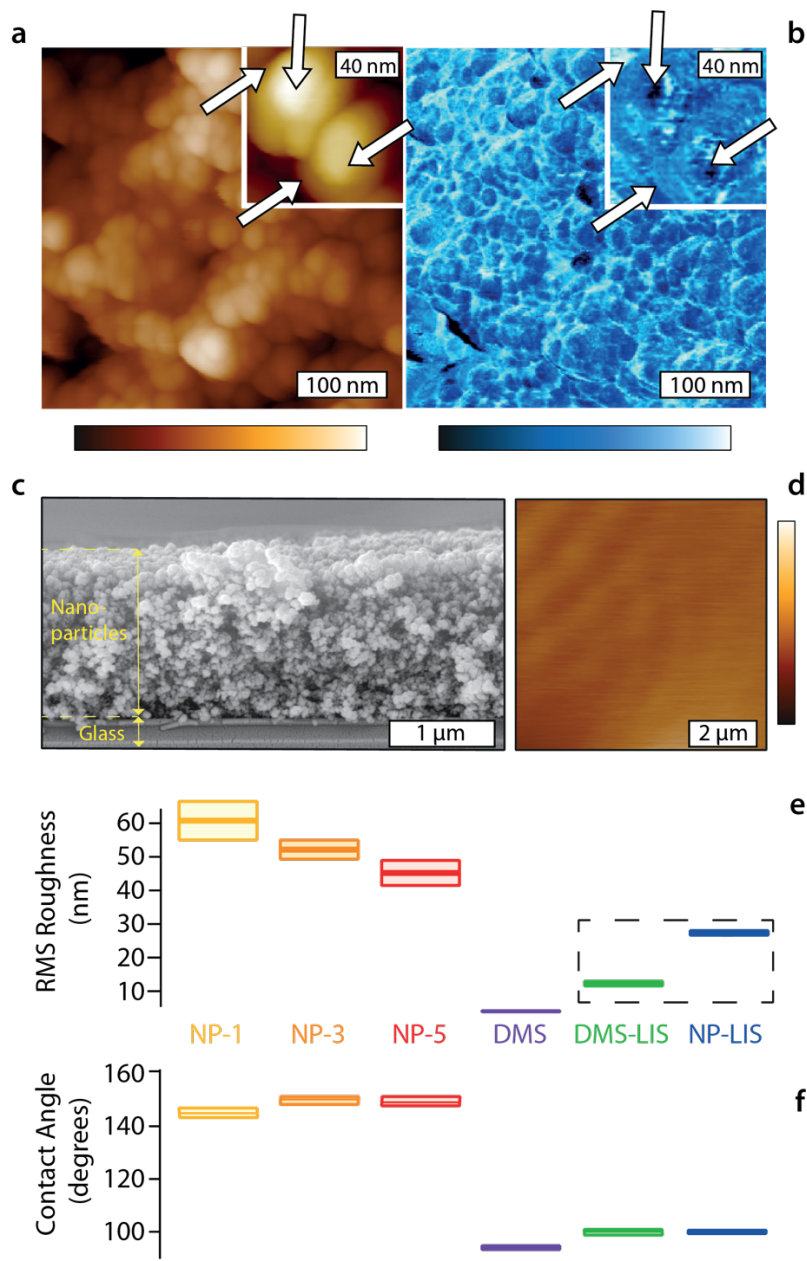

*Figure 1: Characterization of the different surface functionalisations used to create model LIS.* AFM images of the NP-functionalized glass surface taken in ultrapure water (a-b) reveal a cohesive but rough NP layer with multiple NP clusters (a, topography). The associated phase information (b) exhibits some contrast between the harder silica core (darker) and the softer perfluoroalkyltrichlorosilane corona which

*appear lighter and less well defined (inset with arrows indicating the center and edge of two NPs). An SEM image of the section of the NP-functionalized glass taken after 5 NP layers were applied shows a homogeneous ~1.6 µm thick NP coating (c). After infusion with silicone oil, AFM imaging of the oil-water interface shows a smooth surface with occasional tip-induced ripples (d). The surface roughness of the NP-functionalized glass tends to decrease with the number of NP layers sprayed but remains significantly larger than the glass substrate directly coated with a single evaporated layer of dichlorodimethylsilane (DMS) (e). The RMS roughness of each sample was systematically quantified by analyzing 3 distinct regions of on 2µm x 2µm in each case. All the measurements were taken in air except for the LIS, yielding an overall RMS roughness ranging between 20 and 50 nm. Roughness measurements conducted on the liquid infused surfaces are unreliable due to the probing technique inducing ripples in the oil layer. The corresponding roughness values should be considered as indicative only (dashed box). Static CA measurements conducted over all samples (f) yield a value of ~150° for all the NP-functionalized surfaces (within experimental error), changing to 101 ± 1° after infusion with silicone oil. On DMS-functionalized surfaces the CA changes from 95 ± 1° to 101 ± 1° upon infusion with silicone oil. The plotted CA values represent 20 or more independent measurements on a minimum of 2 samples for each set of surfaces. The error boxes (e, f) represent the standard error on the measurements. Samples were stored horizontally and measurements taken over the entire slide length. The color scale bar represents a height variation of 140 nm (a), 60 nm (inset a), 500 nm (d) and phase variation of 30° (b) and 10° (inset b).*

AFM images of a sample taken in ultrapure water, immediately after coating with a single layer of NPs (Fig. 1a-b) reveal a full layer composed of NPs 20-50 nm in size, and creating a surface with an average root-mean-square (RMS) roughness of 60 ± 5 nm. Interestingly, the associated phase image (Fig. 1b) shows additional contrast over single NPs with the cores appearing darker than the hydrophobic shell (inset with arrows). The phase image is sensitive to the local viscoelastic properties of the surface[38,45,46], and able to distinguish between the stiffer silica cores and the softer perfluoroalkyltrichlorosilane shell when operated with a sufficiently large imaging amplitude.[47] Consistently, the cores appear well-defined, but the shells induce a fuzzy outer ring. The NPs are often clustered, which can lead to the solid centers appearing to overlap in some areas, with a range of particle sizes. This can also be seen in the EM image taken on a glass substrate. The layer is uniform and around 1.6 µm thick. The NPs layers were extensively characterized by AFM in air and water, demonstrating them to be consistent, stable and well-attached (Supporting Information Fig. S2). Increasing the number of layers tends to decrease the surface roughness of the coating (Fig. 1e, see also Fig. S2). This is expected, as additional layers allow for newly added particles to settle in more stable positions by filling up grooves in the previous coating layer. After 5 coats, the RMS roughness is comparable to the size of single NPs (~40 nm), suggesting 5 layers to be close to the optimal limit for a smooth, regular coating (45 ± 5 nm). In order to create control surfaces, we used glass slides coated with a single monolayer of dichlorodimethylsilane (DMS) directly evaporated onto the surface, resulting in a significantly lower roughness of 9 ± 1 nm when imaged in identical conditions (Fig. 1e). The DMS-functionalized surfaces can be considered flat and hydrophobic, and hence serve as a reference to single out the effect of porosity on oil retention when compared to a NPs-functionalized surface.

To create full LIS, the NP- and DMS-functionalized glass slides were infused with silicone oil (see Materials and Methods). AFM imaging of the oil-water interface atop an infused 5-layers NPs sample reveals a smooth regular surface with occasional ripples, presumably due to the AFM tip probing the oil-water interface and causing small disturbances (Fig. 1d). Estimates of the average oil layer thickness based on the weight of the samples after infusion yields values between 6 and 8 μm. These values are typical, but can vary, depending on the oil temperature during spin coating, and the environmental conditions in the laboratory.

Static CA measurements, taken across the different surfaces, show the biggest difference between the infused and the non-infused surfaces for any type of functionalization (Fig. 1f). For NPs-functionalized surfaces, infusion with silicone oil reduces the CA from ~150° to ~101°. Within experimental error, no differences can be observed between NP-functionalized surfaces with different numbers of layers which remain stable over days (Fig. S3). This is not surprising, given that we expect the oil to fully cover the surface corrugations. For DMS-functionalized surfaces, the CA changes from ~95° to ~101° upon silicone oil infusion, with an initial measured oil layer thickness comparable to that of the NP-functionalized surfaces (Fig. S4). Overall, the CA values are identical for all fresh LIS within experimental error, reflecting the fact that the CA is then entirely determined by the oil-water interfaces, with no direct effect on the functionalization. The measured values are also in agreement with predictions based on a liquid-equivalent of Young's contact angle equation,[21]

$$\theta = \cos^{-1}\left(\frac{\sigma_{og} - \sigma_{ol}}{\sigma_{lg}^{eff}}\right) = 104 \pm 2° \qquad (1)$$

where $\sigma_{og}$, $\sigma_{ol}$ represent the interfacial energies between oil and air, and oil and water respectively, and $\sigma_{lg}^{eff}$ the effective interfacial energy between water and air. For simplicity, here we have used $\sigma_{lg}^{eff} = \sigma_{og} + \sigma_{ol}$, since the water droplet is expected to be cloaked by the silicone oil. We note that CA values can exhibit changes of several degrees when measured multiple times over days due to changes in the ambient laboratory relative humidity and temperature (Fig. 2). The error associated with the above derivation takes into account such changes which will be discussed later in detail.

Next, to ensure the basic stability of our model LIS, we examined the effect of surface porosity on oil retention as the sample aged unperturbed in air. This was done by comparing the ability of a 5-layer NP-LIS and the control DMS-LIS to retain the infused silicone oil, evaluated by periodic weighing and CA measurements. The results are shown in Figure 2.

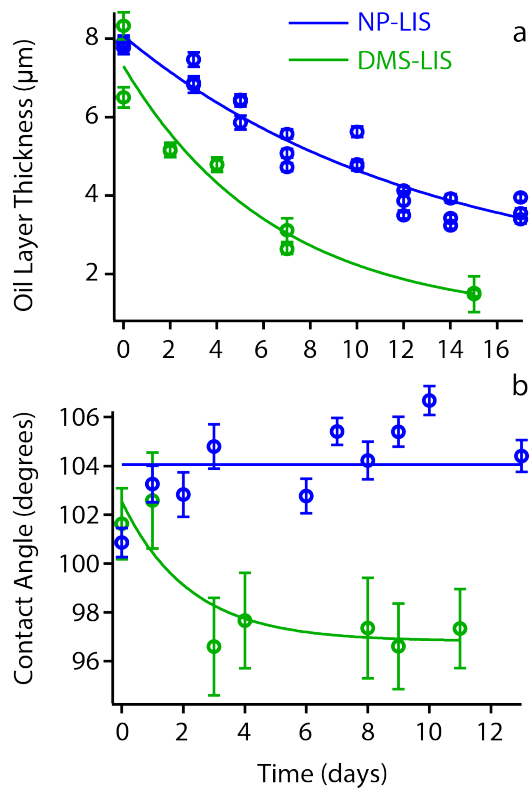

*Figure 2: Oil retention ability of DMS- and 5-layer NP-LIS in air under ambient laboratory conditions compared to the static CA. The thickness of the silicone oil layer decreases in time due to losses to the environment (a). The loss is significantly more pronounced for the DMS-functionalized LIS which depletes to below an oil layer thickness of 2 µm within 15 days. Static CA measurements (b) show no evolution over the NP-functionalized LIS, but a progressive return to oil-free values for the DMS-functionalized LIS. This*

*is consistent with the almost complete oil depletion measured in (a). The two data sets are independent and were taken on different samples but placed together for comparison. Each point represents the average of 3 measurements with its standard error. The solid lines in (a) and (b) serve as eye-guides.*

Static CA measurements on the freshly oil infused LIS give a similar contact angle consistent with Fig. 1a. As the LIS age, oil is lost from both the porous and non-porous surface, but the oil loss occurs more rapidly from the non-porous surface, with the initial oil layer thickness halving over ~4 days. This rapid loss correlates to a change in CA for the DMS-LIS, a behavior not observed for the NP-functionalized LIS, even after partial oil loss. For the NP-LIS, the contact angle remains constant at ~104º. The apparent insensitivity of the CA on the oil layer thickness for the NP-based LIS has been demonstrated by in previous computer simulation studies. These predict a negligible change in CA for relatively large water droplets (> 2-3 mm as used here) on thin oil films upon changes in the film thickness[21]. This is because the typical size ratio between the oil film and the droplet is very small (less than 0.01). We note that this is not true for smaller aqueous droplets whose oil ridge becomes comparable to the droplet size, resulting in an apparent CA that noticeably depends on the oil layer thickness[48].

The main source of error in the experimental measurements of CA comes from fluctuations in the laboratory's temperature $T$ (16 °C < $T$ < 25 °C) and relative humidity $RH$ (50 % < $RH$ < 90 %), both of which are not controlled throughout the experiments. The associated variations in CA over the same sample are ~2% due to $T$ variations and ~2% due to RH variations (See section 4 of the Supporting Information). These uncertainties are in agreement with previous reports indicating CA variations of up to 15% [49]. Changes in $RH$ would, in principle, also affect the droplet evaporation rate, but given the short experimental timescales and the relatively large droplets this can be neglected here. Oil loss over the course of hours can be considered negligible, as can be the impact of gravity due to vertical storage of the LIS (see Supporting Information Figs. S4 and S5 respectively). Overall, the results presented in Figs 1 and 2 confirm the suitability of the 5-layer NP-LIS as a model system to investigate ageing of LIS under external perturbations. The 5-layer NP-LIS will hence be used systematically hereafter unless otherwise specified.

*Accelerated LIS Ageing*

To assess the impact of ageing in more realistic conditions, the 5-layer NPs-LIS samples were immersed in aqueous solutions containing either ultrapure water or a 600 mM NaCl (saline) solution and exposed to pressure waves by ultrasonication. The choice of the saline solution is to mimic conditions in a number of LIS applications, such as for medical devices, transport vehicles and underwater structures. The ageing process used here is harsher than most natural conditions (Table 1) and can be seen as accelerated ageing.

*Table 1: Comparison of the mechanical energy experienced by surfaces in various natural and experimental situations.* *The typical power per surface area associated with natural processes such as the impact of an ocean wave or a raindrop. The values are compared with that calculated for the ultrasonic bath used for accelerated ageing of the LIS in this study. Detailed calculations for the different estimates can be found in sections 1 and 5 of the Supporting Information.*

| Process | Energy [Wm$^{-2}$] |
|---|---|
| Ultrasonic waves (bath sonicator) | 400 – 1200 |
| Ocean wave | 100 – 6000 [50,51] |
| Rain fall (vertically, per drop) | 0.6 – 12 [52] |
| Rain fall (on a windscreen @ 100 km/h, per drop) | 120 – 620 |

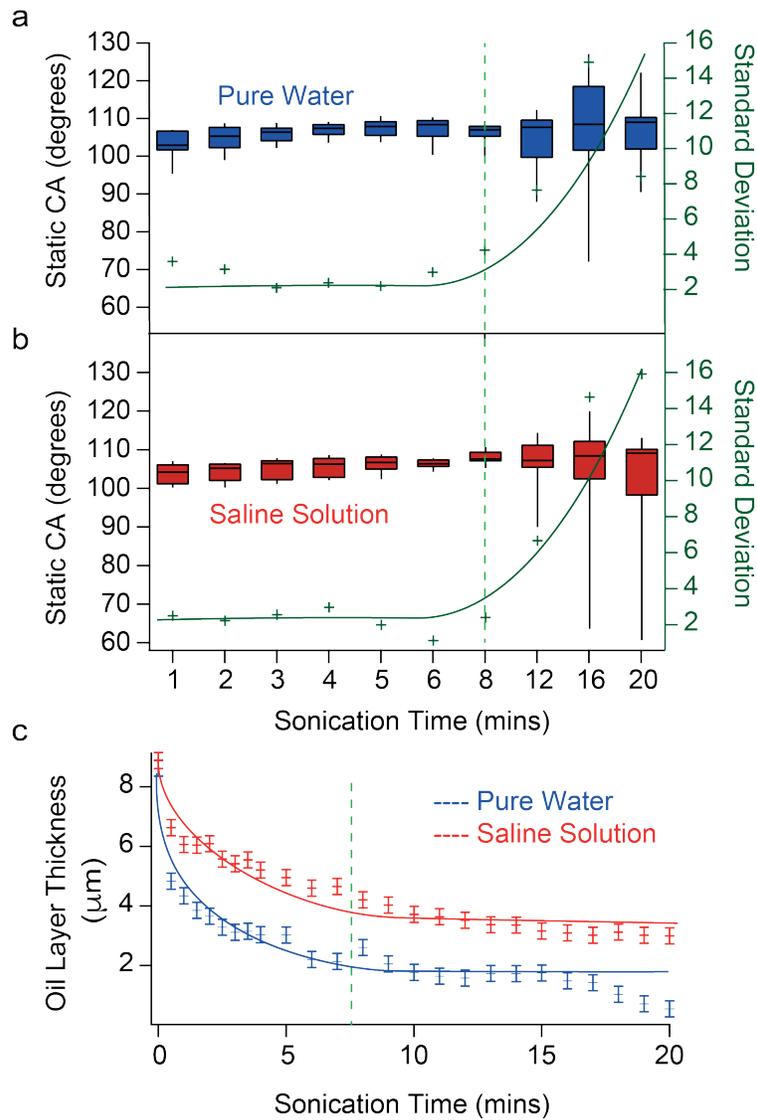

*Figure 3: Accelerated ageing of the model LIS under ultrasonication in ultrapure water and in a 600 mM NaCl (saline) solution.* For both ultrapure water (a) and saline solution (b), the evolution of the CAH is presented in box and whisker plots (black) showing the median value and the upper and lower quartiles. The standard deviation (red) is shown as a function of sonication time with a fitted curve serving as an eye guide. In both liquids, the average CA remains unchanged within error as the infused oil layer thickness decreases (c), but the spread of the CA values increases rapidly past ~8 min of sonication (green dashed line). This indicates significant fluctuations arising with time, presumably due to pinning effects as the oil

*layer progressively becomes patchy. The oil layer thickness (c) was deduced from weight measurements taken every 30s (< 5mins) or every 60s (> 5 mins). Separate samples were used for the CAH data (a, b) and the weight measurements in (c) to avoid the extended time periods necessary to take CAH hysteresis measurements between weight measurements. The data represents 20 CA measurements taken over 5 different locations for each sample and at each time step for the box plots (a, b).*

In both the pure water and the saline solution, the static contact angle remains on average constant over time (Box plot in Fig. 3a, b) with no observable trend within error (standard deviation of the measurements). This is despite an exponential decay in the oil layer's thickness (Fig. 3c). Both solutions exhibit a large spread of measured CA values with a rapid increase past 8 minutes of sonication. This transition approximately coincides with the point where the oil layer thickness starts to plateau after an initial rapid decrease (Fig. 3c). This behavior suggests the appearance of defects in the oil layer, with possible localized exposure of the NP-functionalized surface underneath. This exposure is localized enough not to affect the CA value on average, but sufficient to induce droplet pinning and a higher CA variability. To better quantify this effect and confirm its origins, we conducted contact angle hysteresis (CAH) measurements on a new set of ageing samples, simultaneously tracking the oil layer thickness (Fig. 4). While the strategy allows for less measurement points than in Fig. 3, it provides a more complete picture of the ageing process.

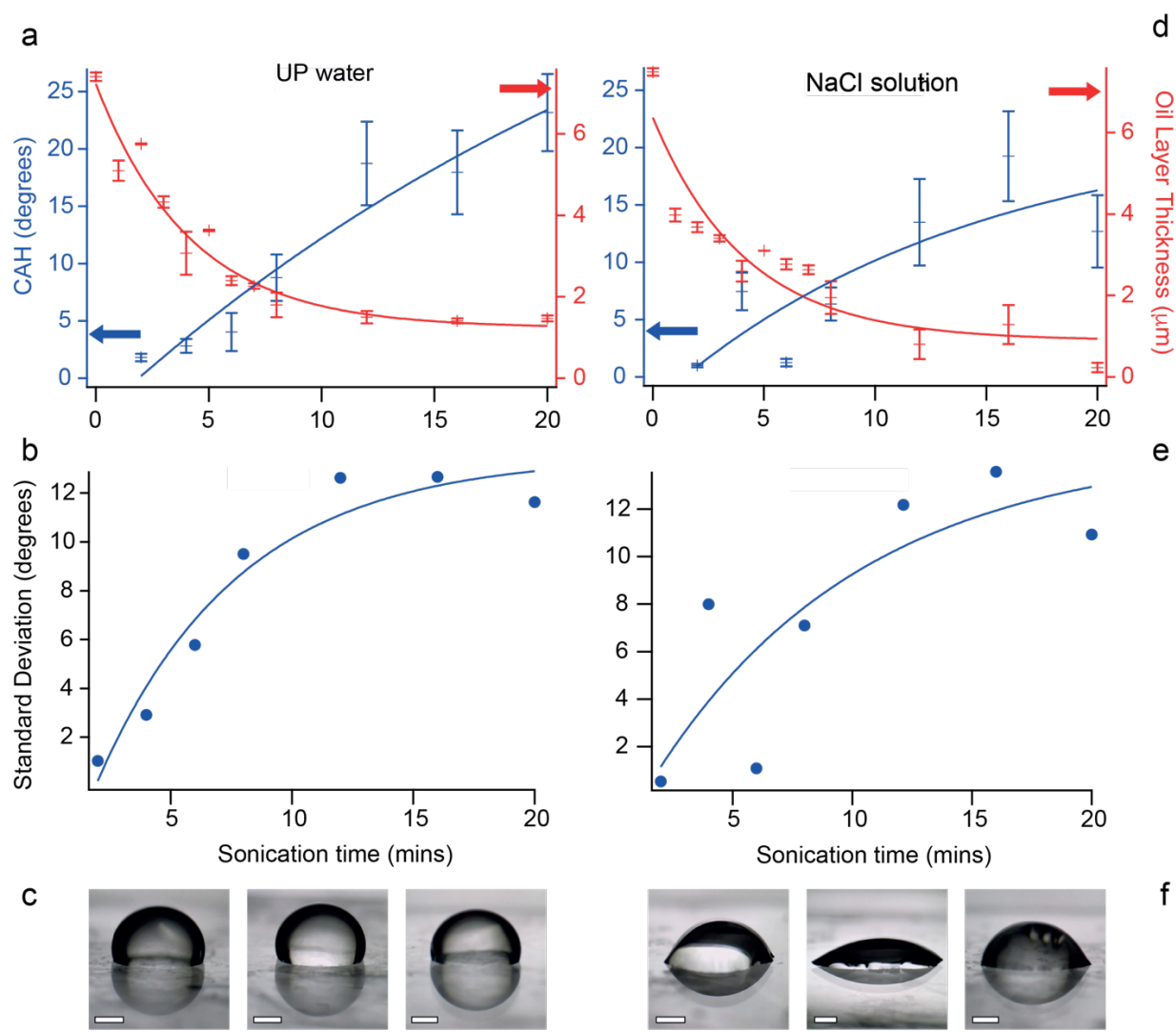

*Figure 4: Changes in CAH upon accelerated ageing of LIS in ultrapure water and a 600 mM NaCl solution.* The evolution of the CAH is shown in both solutions (a, d) with an exponential fit (blue line) and the errors bars representing the data standard error. The infused oil layer thickness is superimposed (red curve, exponential fit) for comparison. The standard deviation of the CAH (b, e) can be seen to increase with the sonication time, supporting the hypothesis of localized defects forming in the infused layer. Images (c) and (f) show representative droplets at early and late sonication times in both media (left to right). Each CAH data point in (a) and (d) represents 12 measurements taken over 4 samples for each solution. Arrows indicate the direction of the reference axis label. The scale bars in images (e) and (f) represent 1 mm.

The CAH values in both solutions are initially low, starting with ~2° and increasing to ~6° with a wider spread after 4 minutes of sonication (Fig. 4a). The hysteresis spread rapidly increases up to values exceeding 30° for times exceeding 12 minutes. This variability can be directly visualized by comparing representative images of the aqueous droplets sitting on the LIS during ageing (Fig. 4e-f): little variation is seen in early droplets, whereas the droplets on aged LIS exhibit pinning and significant variations between droplets, consistent with the apparition of multiple localized defects in the oil layer. The details of these local surface changes are not trivial. Simple oil depletion would be consistent with the increased CA variability and CAH values, but directly exposing hydrophobic NPs should increase the CA upon ageing. This is clearly not the case (Fig. 3) pointing to localized structural and chemical changes to the porous NPs structure.

*Ageing Mechanisms*

Examining the nanoscale details of the ageing porous NPs structure indicate that several related processes are simultaneously operating during the depletion of the initially thick oil layer (Fig. 5). Firstly, the oil removal partly exposes the NPs structures underneath enabling them to become visible by AFM (Fig. 5a, b). Secondly, the exposed NPs-functionalized regions degrade in time, as will be shown later in Fig. 6. (It is worth noting that these exposed features in general have distinct wetting properties compared to surface regions which are never infused.) Thirdly, water microdroplets get trapped in the oil layer (Fig. 5c, d), locally changing the layer's wetting properties, and paving the way for degradation of the NP-functionalized surface. This entrapment of the aqueous solution can be directly visualized at the nanoscale by AFM imaging, with aqueous microdroplets appearing as circular depressions in the oil-water interface (arrows in Fig 5c, d), leading to macroscopic consequences for the LIS' wetting properties (Fig 5e, f). The entrapment results in oil droplets being pinched off the surface, and previously submerged microdroplets leaving the oil layer cloaked. As a result, it is impossible to remove the sonicated LIS from its aqueous bath without losing some of the infused oil layer.

This effect could be confirmed by assessing the stability of pure water and saline emulsion formed in silicone oil by sonication (Fig. S6). Over a timescale of hours, the microdroplets can been seen, on average, to increase in size with the saline solution exhibiting a slightly higher stability. This is likely to be due to the fact that saline droplets have a higher colloidal stability than their pure water counterpart where only hydroxyl ions are able to stabilize the droplet.[53] When NaCl is added at high molarity, the co-ions preferentially sit at the oil-water interface next to the hydroxyl ions, effectively creating an ionic surfactant layer which renders the droplet positively charged and more stable against coalescence.[54]

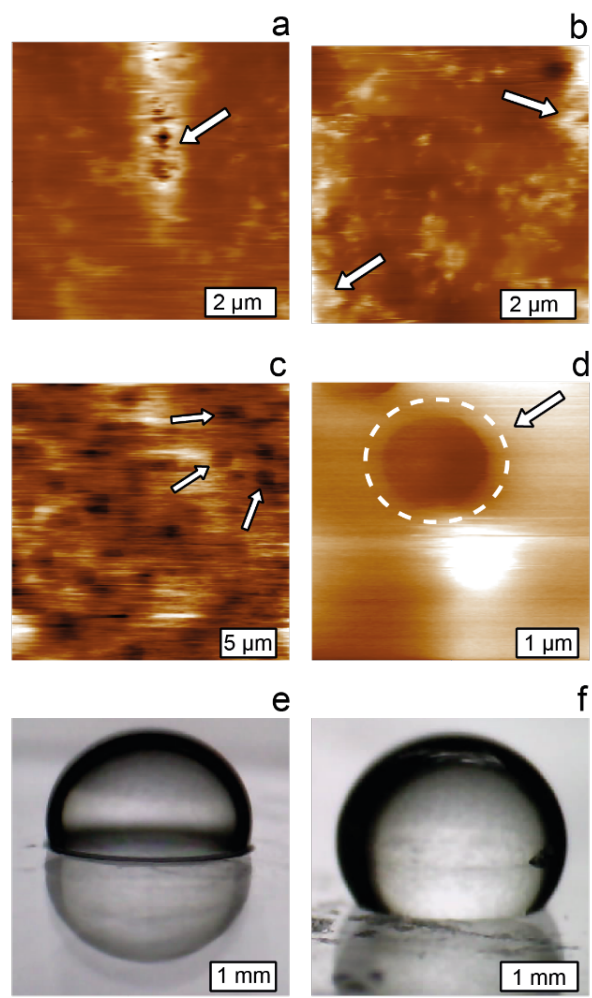

*Figure 5: Images of the oil-aqueous solution interface during accelerated ageing.* AFM images reveal a progressive depletion of the infused oil-layer, exposing some surface features of the NPs nanoporous layer (a). More features are visible at longer sonication time (b), consistent with the oil layer shedding to reveal the surface underneath with small nano-ridges emerging from the depleted oil layer (arrows in a, b). The main mechanism for oil removal (c) is the replacement of oil microdroplet by aqueous droplets, inducing characteristic circular depleted regions in the oil layer (arrows). A magnified view of one such circular depletion is highlighted by a white dashed circle and arrow (d). The horizontal streaks in (c-d) confirm that the AFM tip is still scanning a fluid and mobile layer. Optical images of droplets on fresh (e) samples show an oil ridge at the drop edge. When the layer is sufficiently depleted (f), the oil ridge is no longer visible.

*The color scale is as for Fig. 1, with a maximum height variation of 124 nm (a), 242 nm (b), 129 nm (c) and 90 nm (d).*

The porous NPs-layer substrate used for the LIS ages too as a result of sonication. This occurs not only as a direct, mechanical result of the sonication waves, but also by increased exposure to the aqueous solution[29], an effect exacerbated in the presence of more stable saline droplets. Control sonication experiments carried out on NPs-functionalized surfaces without any oil present show that direct mechanical effects mainly decrease the porous layer's roughness from ~60 nm to 35-40 nm (Fig. S7). This is likely due the removal of loose or protruding particles, leaving a more uniform surface. As can be expected, the wetting properties of the surface change with the roughness[55,56] (Fig 6a; NP-5 in water). Superhydrophobic surfaces have poor underwater stability due to the difficulty in retaining air pockets[29,57]. Here, sonication could force water against the surface, displacing any residual air pockets. However, the surface itself remains fully and uniformly covered with the NPs well attached (Fig. S7). This suggests a limited impact of the pressure waves on the integrity of the nanoporous surface. Instead, damage to the nanoporous layer mainly results from prolonged exposure to the saline solution, which can cause the NPs to detach or become loose.

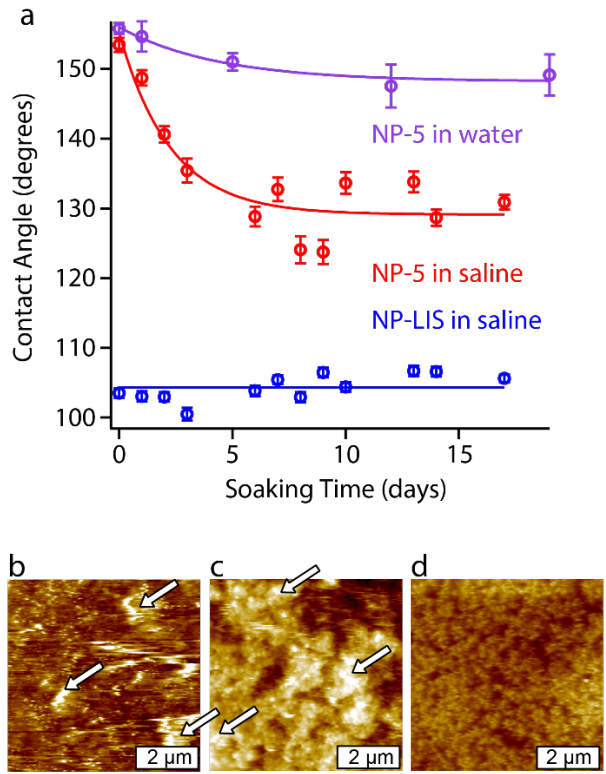

*Figure 6: Impact of NP-5 soaking in an aqueous solution on the integrity and properties of the NPs layer. Over 20 days, the static CA decreases significantly on the nanoporous NP-5 surfaces in saline solution while no evolution is seen for the LIS (a). AFM images taken on the NP-5 surface at day 5 in pure water (b), the NP-5 surface at day 2 in saline solution (c), and the LIS surface at day 14 in the saline solution (d), reveal some key differences in ageing. Permanent surface degradation is evident for the non-infused surfaces (b, c) where large clusters are present (arrows) and more pronounced in the saline solution. Increased roughness is also visible on the LIS but much less pronounced, and the characteristic scanning streaks confirms the presence of a mobile oil layer. The color scale in all AFM images represents a height variation of 400 nm (as for Fig. 1).*

Upon soaking in a saline solution, the un-coated nanoporous NPs layer exhibits a progressive decrease of static CA values (Fig. 6a; NP-5 in saline), suggesting a transition between two wetting states, similar to a Cassie to Wenzel transition. This interpretation is supported by the presence of some degradation of the surface with localized irregularities and NPs clustering, revealed by AFM (Fig. 6c). The surface is also fragile with NPs easily removed during AFM imaging after only two days of soaking. The contact angle measurement (5 layers NP-LIS in saline, Fig. 6a) does not show any significant change. This result indicates that without any external perturbation the oil layer provides a protective coating, preventing the solution from interacting with the NPs and degrading the surface (Fig. 6d). If the solution comes into contact with the NPs, it can destabilize and modify the surface, likely due to metal ions facilitating the removal of the hydrophobic ligands from the surface of the silica NPs. Since the ligand is tethered to the silica core by silane chemistry, they can be displaced under appropriate conditions[58]. This would result in hydrophobic ligand clusters aggregating on the surface, consistent with the AFM images of the degraded surface (Fig. 6b, c).

*General discussion*

LIS have the potential to revolutionize antifouling coating, offering a more efficient and environmentally friendly alternative to existing solutions. However, any real-life application requires a clear understanding of the LIS ageing over time so as to enable the targeted development of better, more flexible surfaces that can withstand the demands of their intended application. Here, we track the ageing of model LIS prepared according to standard protocols. Using a dual micro- and macroscale experimental strategy, we link functional changes in the LIS' performance with specific oil loss mechanisms and nanoscale effects in the porous layer. We find that the initial oil layer is usually not at equilibrium, leading to significant oil loss, even when stored in ambient conditions and without any external perturbation. When immersed in aqueous solutions and exposed to high intensity ultrasonic pressure waves, the oil loss significantly accelerates, inducing changes in contact angle and contact angle hysteresis. The pressure waves create aqueous microdroplets in the oil layer, progressively pinching out the oil from the LIS, as cloaked droplets move out of the oil phase back into the bulk aqueous solution. The rapidity of the depletion process is weakly dependent on the colloidal stability of the microdroplets. This mechanism, to the best of our knowledge not previously reported, appears central to the ageing of LIS exposed to the impact of waves. In our simple model LIS, this 'pinching' mechanism induces an important side-effect: the irreversible degradation of the hydrophobic NP coating used to retain the LIS' liquid. The NP coating plays an important role in the LIS performance which is maintained more than ten times longer for the nanoporous NPs-coated supports compared to the smooth flat supports. However, the current findings highlight inherent weaknesses in the use facile nano-structured NP coatings exposure of the LIS support to the aqueous solution causing both chemical and structural degradations.

Taken together, the present results provide clues to designing robust LIS, for example aimed at real-life applications that entail the impact of water drops or waves. First, using a retention support that does not require chemical functionalization, unlike the hydrophobized particles used here, would offer an obvious strategy to remedy LIS degradation, potentially increasing the lifetime of both the support and the resulting LIS. This is especially true for substrates exposed to saline solutions, where the contaminate degrades the substrate more quickly. If the LIS is designed for being reinfused periodically, chemical resistance to the environment (other than the infusing liquid) is necessary to avoid degradation over short timescales. Second, mechanical restructuring of the porous layer may also need to be considered depending on the application.

# Conclusion

Overall, this work shows that the ageing effect on LIS surfaces can be significant when exposed to non-ideal environmental conditions. Practical and application-oriented developments of LIS are likely to become a key aspect to LIS adoption in technology and industry, beyond the many fundamental developments currently driving the field.[3,9,10,59,60] The ageing mechanisms depend on the specific details of the system considered and should be tailored for the applications of interest.[4–7,28] The present work could act as a reference point for future work involving the testing of new LIS applications, in particular the development of a standardized, accelerated ageing strategy, to determine the robustness and durability of novel products.

# Associated Content
**Supporting information**

Supporting Information Available:

S1: Estimation of the ultrasonic bath characteristics

S2: Calculation of the spin-coated oil layer thickness

S3: Supplementary Results

S4: Effect of humidity and temperature on CA measurements

S5: Estimation of the energy associated with real-life phenomena

S6: Supplementary references

This material is available free of charge via the Internet at http://pubs.acs.org.

# Author Information

**Corresponding Authors**

* Email: halim.kusumaatmaja@durham.ac.uk

† Email: kislon.voitchovsky@durham.ac.uk

**ORCID**

Sarah Goodband: 0000-0003-0378-9310

Steven Armstrong: 0000-0002-0520-8498

Halim Kusumaatmaja: 0000-0002-3392-9479

Kislon Voitchovsky: 0000-0001-7760-4732

**Author Contributions**

SJG, HK and KV designed the experiments. SJG performed all the measurements with help from SA. SJG analyzed the results with help from HK and KV. SJG, HK and KV wrote the paper.

**Funding Sources**

EPSRC for funding through the SOFI (Soft Matter and Functional Interfaces) center for doctoral training (grant EP/L015536/1).

**Notes**

The authors declare no competing financial interest.

# Acknowledgements
The authors would like to thank Lisong Yang and Colin Bain for help with DMS sample preparation (Silanization protocol and use), Bethany V. Orme, Gary Wells and Glen McHale for their help with the preparation of the LIS samples and SEM image and Clodomiro Cafolla for his help with AFM. SJG is grateful to the EPSRC for funding through the Soft Matter and Functional Interfaces (SOFI) doctoral training school.